\documentclass[a4paper]{article}

\usepackage{INTERSPEECH2022}
\usepackage{bbm}
\usepackage{xcolor}
\usepackage{cite}
\usepackage{diagbox}

\title{Self-Supervised Speaker Verification Using Dynamic Loss-Gate and \\ Label Correction}
\name{Bing Han, Zhengyang Chen, Yanmin Qian\textsuperscript{\dag} 
\thanks{\textsuperscript{\dag}Yanmin Qian is the corresponding author}}
\address{
MoE Key Lab of Artificial Intelligence, AI Institute \\
X-LANCE Lab, Department of Computer Science and Engineering \\
Shanghai Jiao Tong University, Shanghai, China
}
\email{\{hanbing97, zhengyang.chen, yanminqian\}@sjtu.edu.cn}

\begin{document}

\maketitle
\begin{abstract}


For self-supervised speaker verification, the quality of pseudo labels decides the upper bound of the system due to the massive unreliable labels. 
In this work, we propose dynamic loss-gate and label correction (DLG-LC) to alleviate the performance degradation caused by unreliable estimated labels. In DLG, we adopt Gaussian Mixture Model (GMM) to dynamically model the loss distribution and use the estimated GMM to distinguish the reliable and unreliable labels automatically.
Besides, to better utilize the unreliable data instead of dropping them directly, we correct the unreliable label with model predictions. 
Moreover, we apply the negative-pairs-free DINO framework in our experiments for further improvement. 
Compared to the best-known speaker verification system with self-supervised learning, our proposed DLG-LC converges faster and achieves $11.45\%$, $18.35\%$ and $15.16\%$ relative improvement on Vox-O, Vox-E and Vox-H trials of Voxceleb1 evaluation dataset.


\end{abstract}

\section{Introduction}

Recently, the deep learning based methods have been broadly applied for speaker verification (SV) task and obtained excellent performance~\cite{xvector, rvector, svector2, partial_share}. 
However, all these methods require large amounts of training data with speaker labels, while the collection of data with annotated speaker labels is often very difficult and expensive. 

In order to make full use of a large quantity of unlabeled data, many efforts~\cite{nagrani2020disentangled, chung2020seeing, inoue2020semi, huh2020augmentation, zhang2021contrastive, xia2021self, mun2020unsupervised} have been made to obtain good speaker representations in a self-supervised learning manner. Following the iterative framework proposed in~\cite{cai2021iterative}, the current state-of-the-art self-supervised speaker verification systems usually include two stages. 
In stage I, a speaker encoder is trained by contrastive learning based loss~\cite{zhang2021contrastive}. 
In stage II, estimating pseudo labels from the pre-trained model and then training a new model based on the estimated pseudo labels are iteratively performed to continuously improve the performance.

This stage-wise training framework has achieved excellent performance, while there still exist some unreasonable assumptions, which limits the further improvements of system performance.
In stage I, contrastive learning based methods assume that speech segments from the different utterances belong to different speakers. When training the network, they will be regarded as negative pairs and be pushed away from each other in the embedding space. There is no doubt that this inaccurate assumption will bring some mistakes, because different utterances may come from the same speaker. 
In stage II, a clustering algorithm is adopted to generate the pseudo labels for each utterance based on the previous learned representation and the estimated pseudo labels will be used in the subsequent supervised training. However, \cite{caron2018deep} shows that there are massive amount of unreliable pseudo labels.
Such low-quality labels will confuse and degrade the model during the training, which indicates that finding a way to select high-quality labels is the key to improving the performance \cite{rizve2021defense}. Cai \textit{etal.} \cite{cai2021iterative} applied an aggressive method to purify the pseudo labels based on clustering confidence, but the improvement is minor.
Besides, Tao \textit{etal.} \cite{tao2021self} observed that the data with lower loss is more reliable than those with unreliable labels. Based on this hypothesis,
they set a loss threshold in each iteration to distinguish the reliable labels from unreliable labels and only use the samples with reliable labels to update the network. Although this approach brings further improvement, there is still a lot of room for improvement. The manually set threshold in each iteration is not flexible and the data with unreliable labels is not fully utilized.

To tackle these problems and improve the existing methods, in this work, we introduce DINO (distillation with no labels)~\cite{dino} in stage I, 
which is only based on maximizing the similarity between feature distributions of differently augmented segments from the same utterance so it's free from the negative pair issue. 
Our experiments show that DINO outperforms all previous contrastive learning based methods in Voxceleb dataset~\cite{voxceleb1,voxceleb2}. In stage II, we model the loss distribution of all the data using Gaussian Mixture Model (GMM), more specifically, GMM with two components, in which each Gaussian component can represent the loss distribution of the data with reliable labels or the data with unreliable labels. 
With the estimated GMM, we can automatically obtain the loss threshold to distinguish the two types of labels, which is more flexible than the manually tuned threshold in~\cite{tao2021self}.
Besides, in many semi-supervised learning works~\cite{berthelot2019remixmatch,sohn2020fixmatch,zhang2021flexmatch}, a pre-trained model is often used to label an unlabeled dataset.
Similarly, instead of directly discarding the data with unreliable labels in~\cite{tao2021self}, we leverage the model prediction as an alternate label and use it to correct the unreliable label.
We named the above two strategies that we proposed in stage II as \textit{dynamic loss-gate and label correction} (DLG-LC). With DLG-LC,
we build a more robust self-supervised speaker verification system and obtain an equal error rate of 1.468\% on  Voxceleb 1 test set which is the \textbf{state-of-the-art} nowadays.

\begin{figure*}[t]
  \centering
  \includegraphics[width=0.8\linewidth]{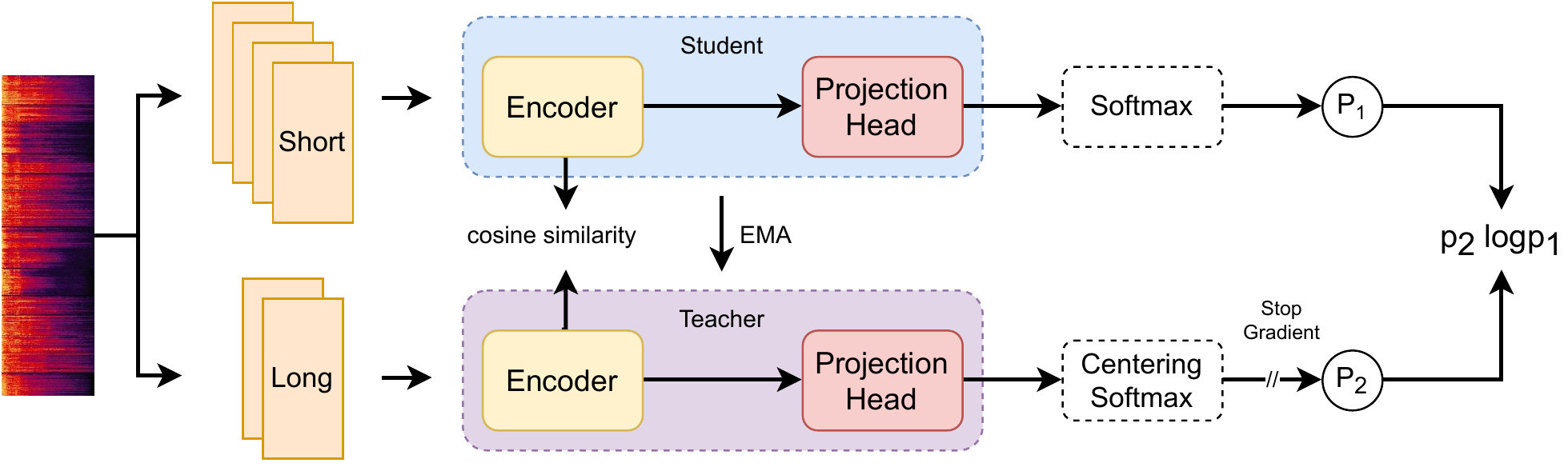}
  \caption{Framework of Distillation with no label (DINO) for self-supervised speaker representation learning}
  \label{fig:dino}
\end{figure*}

\section{Methods}
\subsection{DINO based Self-Supervised Learning}

Inspired by~\cite{dino}, we introduce DINO which doesn't need negative pairs, and the whole framework is shown in Fig.~\ref{fig:dino}. 
Firstly, we sample 4 short segments $\{x_1^s \dots x_4^s\}$ and 2 long segments $\{x_1^l,x_2^l\}$ from a utterance with a multi-crop strategy~\cite{caron2020unsupervised}, where the long segments can be used to extract more stable speaker representation. 
We assume that the segments cropped from the same utterance share the same speaker identity and then augment them in different ways by adding noise or room impulse response for robust performance. In the following, all short segments are passed through the student while only the long segments are passed through the teacher, therefore encouraging \textit{short-to-long} correspondences by minimizing the cross-entropy $H(\cdot)$ between two distributions:
\begin{equation}
\label{equ:ce}
L_{ce} = \sum_{x\in \{x_1^l,x_2^l\}}^{} \sum_{x'\in \{x_1^s \dots x_4^s\}}^{} H(P_t(x) \mid P_s(x')) 
\end{equation}
where $P_t$ and $P_s$ represent the output probability distributions of teacher network $g_{\theta_t}$ and student network $g_{\theta_s}$ respectively, and we can compute $P$ by normalizing the output with a softmax function:
\begin{equation}
\label{equ:sharpen}
P_s = Softmax(g_{\theta_s}(x)/\epsilon_s )
\end{equation}
with $\epsilon_s>0$ a temperature parameter that controls the sharpness of the output distribution, and a similar formula holds for $P_t$ with temperature $\epsilon_t$. In addition, the output of the teacher network is centered with a mean computed over the batch. Sharpening and centering are both adopted for avoiding trivial solution~\cite{dino}.

Both teacher and student have the same architecture but different parameters. The student is updated by gradient descent while the teacher is updated by an exponential moving average (EMA) of the student's parameters. The update rule is $\theta_t \leftarrow \lambda \theta_t + (1-\lambda) \theta_s$, with $\lambda$ following a cosine schedule from $0.996$ to $1$ during training. We extract speaker embeddings by Encoder and feed them into Projection Head which consists of a 3-layer MLP with hidden dimension 2048 followed by $\ell_2$ normalization and a weight normalized fully connected layer with \textit{K} dimensions, whose architecture is similar to~\cite{caron2020unsupervised}. 

In order to make DINO more suitable for speaker verification, 
we also add a consistency loss to maximize the cosine similarity between the embeddings from the same utterance. This loss is design to ensure that the speaker is embeded in cosine space which is more suitable for scoring and clustering in the following. Then, the final loss is jointed by coefficient $\alpha$:
\begin{equation}
    L_{dino} = L_{ce} + \alpha \sum_{e\in \{e_1^l,e_2^l\}}\sum_{e'\in \{e_1^s \dots e_4^s\}} ( 1 - \frac{e \cdot e'}{\left \| e \right \| \left \| e' \right \| })
\end{equation}
where $e$ denotes the speaker embedding extracted by encoder.

\subsection{Dynamic Loss-Gate}
Following the iteration framework proposed in~\cite{cai2021iterative}, we use the speaker encoder pre-trained by DINO to extract the speaker embeddings for each utterance. Then, we apply $k$-means clustering algorithm to assign the same pseudo labels to the utterances which belong to the same cluster. Based on these pseudo labels, a new speaker encoder can be trained to generate more robust speaker embeddings. This process will be repeated several iterations until the model converges. 
In this process, data with unreliable labels will undoubtedly degrade the performance of the model.
In~\cite{tao2021self}, they observed a distinct separation between the loss of the data with reliable and unreliable labels in the final converged stage and proposed a loss-gated learning (LGL) method to effectively select the reliable data using a fixed threshold. With LGL, it has achieved amazing results, but the threshold setting of this method is very dependent on human experience, and the unreliable data are not fully utilized.

\begin{figure}[ht]
  \centering
  \includegraphics[width=0.8\linewidth]{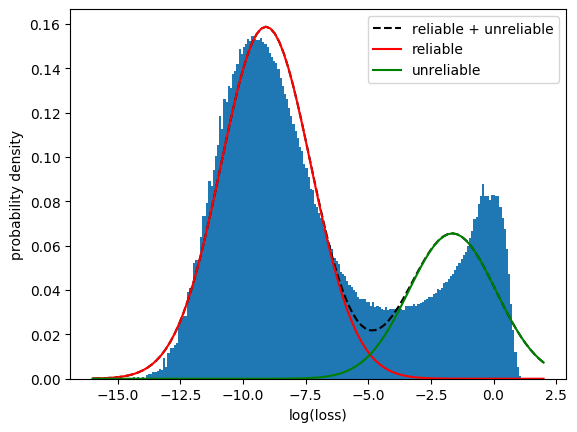}
  \caption{Loss distribution of LGL on Voxceleb 2. Loss value is scaled by log function. And the lines are estimated by GMM.}
  \label{fig:loss_distribution}
  \vspace{-0.5cm}
\end{figure}

In order to obtain an appropriate threshold, we implement the LGL and conduct an experiment to analyze the loss distribution on Voxceleb 2~\cite{voxceleb2}. The histogram is shown in Figure \ref{fig:loss_distribution}. From the loss distribution, it is obvious to find that the distribution has two sharp peaks. Similar experiments in \cite{arazo2019unsupervised} have shown that these two peaks can represent the data with reliable and unreliable labels respectively.
If we can find a way to model this distribution, then
the threshold can be calculated directly and dynamically varies with the loss distribution, which can also avoid laborious manual tuning.

Gaussian distribution is an important type of continuous probability distribution for a real-valued random variable. And the general form of its probability density function is in Eq.\ref{equ:gaussian}.
\begin{equation}
\label{equ:gaussian}
\mathcal{N}\left(\mu, \sigma^{2}\right) = \frac{1}{\sigma\sqrt{2\pi}} exp(-\frac{1}{2} (\frac{x-\mu}{\sigma})^2)
\end{equation}
where $\mu$ is the location parameter and $\sigma$ is the scale parameter. It is bell-shaped, with low on both sides and high in the middle, whose shape is very similar to "peaks" of loss in Figure~\ref{fig:loss_distribution}. In this case, Gaussian Mixture Model (GMM) with 2 components can be employed to fit the distribution of reliable and unreliable samples respectively:
\begin{equation}
\label{equ:2_component_gmm}
p(x)=\lambda_1 \mathcal{N}\left(\mu_1, \sigma_2^{2}\right) + \lambda_2 \mathcal{N}\left(\mu_2, \sigma_2^{2}\right)
\end{equation}
where $\lambda$ is the weight for each Gaussian component.
We plot the fitted curves in Figure~\ref{fig:loss_distribution}, it's obvious that the two ``peaks" can be approached by two weighted Gaussian distributions. Then we can easily obtain a threshold $\tau$ to distinguish the reliable and unreliable data by finding a loss value that the probabilities belonging to both components are equal:
\begin{equation}
\label{equ:obtain_threshold}
\tau : p_1(\tau) = p_2(\tau)
\end{equation}
where $p_1(x) = \lambda_1 \mathcal{N}\left(\mu_1, \sigma_1^{2}\right)$ and $p_2(x)=\lambda_2 \mathcal{N}\left(\mu_2, \sigma_2^{2}\right)$. For each epoch, we will re-estimate the GMM based on the recorded loss values in that epoch, so $\tau$ can be adjusted dynamically according to the current training condition.

Similar to the LGL, we introduce this threshold $\tau$ into the speaker classification loss function ArcMargin Softmax (AAM)~\cite{aam} to filter the data with small loss and only use the retained data to update the parameters of the network.
\begin{equation}
\label{equ:reliable_prob}
    L_{DLG} = \sum_{i=1}^{N} \mathbbm{1}_{l_i < \tau}  \log\frac{e^{s(\cos(\theta_{y_i,i}+m))}}{Z}
\end{equation}
where $Z=e^{s(\cos(\theta_{y_i,i}+m))} + \sum_{j=1,j\neq i}^{c} e^{s(\cos(\theta_{y_i,i}))}$, $\theta_{j,i}$ is the angle between the column vector $W_j$ and embedding $x_i$. $s$ is the scaling factor and $m$ is  hyperparameter to control the margin. 
AAM can enforce larger gaps between the nearest speakers and is widely used in speaker recognition tasks.

\subsection{Label Correction}
To effectively utilize the unreliable data with large losses instead of directly throwing them away, we also propose label correction (LC) to correct pseudo labels dynamically. 
As the model is continuously optimized, the output of the model for unreliable samples can reflect their true labels to some extent~\cite{tong21_interspeech}.
To leverage this ability, we assume that the model's output prediction is more reliable than pseudo labels which are generated by clustering. Therefore, we incorporate the predicted posterior probability, as the target labels, into the objective function to prevent the model from fitting into incorrect labels.
However, not all prediction labels are suitable for training. Inspired by~\cite{lee2013pseudo, sohn2020fixmatch}, we assume that a prediction label owns high confidence if the model assigns a high probability to one of the possible classes. Therefore, we introduce another threshold $\tau_2$ to retain prediction labels whose largest class probability is above $\tau_2$. And the label correction loss is defined as the following:
\begin{equation}
    L_{LC} = \sum_{i=1}^{N} \mathbbm{1}_{l_i > \tau, \max(\hat{p_i})>\tau_2} H(\hat{p_i} \mid p_i)
\end{equation}
where $p_i$ and $\hat{p_i}$ represent the output probability of augmented segments and their corresponding clean version respectively. $H(\cdot)$ denotes the cross-entropy between two probability distributions. We also apply a sharpening operation on $\hat{p_i}$ to encourage a peaky distribution with $\epsilon_c$ which is defined in  Eq.~\ref{equ:sharpen}. 

Finally, these two losses are combined to optimize the speaker model.
\begin{equation}
\label{equ:total_loss}
    L = L_{DLG} + L_{LC}
\end{equation}

\section{Experiment}
\label{sec:exp}

\subsection{Dataset and Configuration}

In our experiments, we adopt the development set of Voxceleb2~\cite{voxceleb2} for training the networks and no speaker label is used during this process. It comprises 1,092,009 utterances among 5,994 speakers. For the evaluation, we report the experimental results on 3 trial sets as defined in [17]: the Original, Extended, and Hard Voxceleb test sets.
The main metrics adopted in this paper are equal error rate (EER) and normalized minimum detection cost function (minDCF).

For both two stages in our experiments, we use 80-dimensional filter bank with 25ms windows and 10ms shift as the acoustic features. In addition, online data augmentation~\cite{online_aug} is applied with noisy MUSAN~\cite{musan} and RIRs~\cite{rir} datasets. 

\subsubsection{Stage I : DINO}
For DINO, considering the training time and memory limitation, we adopt a thin version ResNet~\cite{resnet} to learn speaker representation from the long (3 seconds) and short (2 seconds) segments, and then they are encoded into 256-dimensional speaker embedding by networks. Similar to the configuration in~\cite{dino}, the K in the DINO projection head is set 65536. Temperatures for the teacher $\epsilon_t$ and the student $\epsilon_s$ are 0.04 and 0.1 respectively. In addition, we set $\alpha$ to $0.001$ to balance two losses during the training process.

\subsubsection{Stage II : Iterative Pseudo Label Learning}
In this stage, to have a better comparison with the results in~\cite{tao2021self}, we use ECAPA-TDNN~\cite{ecapa} as our encoder. For clustering, we adopt the \textit{k}-means algorithm to assign the pseudo label to the training set which is supported by faiss library~\cite{faiss}. To verify the robustness of our method, we choose $7500$ instead of $6000$~\cite{cai2021iterative} as the cluster number. When training the network on pseudo labels with DLG, we set the margin as $0.2$, scale as $32$ in the AAM. For LC, we set sharpen parameters $\epsilon_c$ as $0.1$ and threshold $\tau_2$ as $0.5$. The learning rate decays from $0.1$ to $5e$-$5$ exponentially with SGD optimizer. Our training process repeats 5 iterations. In the last iteration, we extend the channel size of the encoder (ECAPA-TDNN) to 1024 following \cite{tao2021self}.

\subsection{Comparison of DINO with Previous Work}
To throughly evaluate the DINO, we conduct experiments and corresponding results are listed in Table~\ref{tab:dino_result}. All the methods are trained on Voxceleb 2 and evaluated on Vox-O test set. From the table, we can find that negative-pairs-free DINO outperforms all previous contrastive-based methods, which show that negative pairs are indeed a bottleneck for performance improvement. In addition, we also provide the ablation study of DINO at the bottom of Table~\ref{tab:dino_result}. We can see that DINO without EMA obtains a very bad result which shows that EMA is the key to preventing the model from collapse. When we add multi-crop and cosine strategies during training respectively, both of them can improve the DINO performance in different degrees and demonstrate their necessity.

\begin{table}[h]
  \caption{Comparison of DINO with other self-supervised speaker verification work. EER (\%) and minDCF are evaluated on Vox-O test set.}
  \label{tab:dino_result}
  \centering
  \begin{tabular}{lrr}
    \toprule
    Method &  EER (\%)  & minDCF \\
    \midrule
    Disent~\cite{nagrani2020disentangled} & 22.09 & - \\
    CDDL~\cite{chung2020seeing} & 17.52 & -\\
    GCL~\cite{inoue2020semi} & 15.26 & -\\
    i-vector~\cite{huh2020augmentation} & 15.28 & 0.63 (p=0.05) \\
    AP + AAT~\cite{huh2020augmentation} & 8.65 & 0.45 (p=0.05) \\
    SimCLR + uniform ~\cite{zhang2021contrastive} & 8.28 & 0.610 \\
    MoCo + WavAug~\cite{xia2021self} & 8.23 & 0.590 \\
    Unif+CEL~\cite{mun2020unsupervised} & 8.01 & - \\
    \midrule
    DINO & 	31.23 & 0.990 \\
    \hspace{1em} + EMA & 7.02 & 0.579 \\
    \hspace{1em} + + Multi-Crop & 6.35 & 0.566 \\
    \hspace{1em} + + + Cosine loss & \textbf{6.16} & \textbf{0.524} \\
    \bottomrule
  \end{tabular}
\end{table}



\subsection{Dynamic Loss-gate with Label Correction}

\begin{table}[h]
  \caption{EER (\%) comparison on Vox-O, E, H of the proposed DLG-LC in Iteration 1. In this experiment, pseudo labels are estimated from our pre-trained DINO system. DINO means we simply used all the data with the estimated pseudo labels as the supervisory signal without any loss-gate during system training.}
  \label{tab:ablation_DLGLC}
  \centering
  \begin{tabular}{lccc}
    \toprule
    Method & Vox-O & Vox-E & Vox-H  \\
    \midrule
    DINO & 4.255 & 4.900 & 8.005 \\
    \hspace{1em} + LGL~\cite{tao2021self} & 3.590 & 4.373 & 6.935 \\
    \hspace{1em} + DLG & 3.202 & 3.525 & 5.805 \\
    \hspace{1em} ++ LC & \textbf{2.723} & \textbf{3.179} & \textbf{5.442} \\
    \bottomrule
  \end{tabular}
\end{table}

Based on pseudo labels generated by pre-trained DINO in stage I, we conducted some experiments to illustrate the effectiveness of our proposed method. Firstly, we give an ablation study of DLG-LC in Iteration 1. From Table~\ref{tab:ablation_DLGLC}, it can be observed that the LGL can bring significant improvement compared with the baseline which is trained without any data selection. It means that loss-gate can effectively filter the reliable labels which are of benefit to the model. However, the choice of threshold also has a non-negligible impact on model performance~\cite{tao2021self}. Based on the estimated GMM, our proposed DLG can dynamically adjust the threshold according to the current training situation and  obtains better performance than LGL which adopt a fixed threshold during the whole training process. In addition, we add the LC strategy to make full use of data with unreliable labels and the results are further improved.  

\begin{table}[h]
  \caption{EER (\%) comparison on Vox-O test set for different iterations of the proposed DLG-LC with other strategies. SimCLR and DINO mean that we simply used all the estimated pseudo labels without any loss-gate during training process.}
  \label{tab:DLGLC_iteration}
  \centering
  \begin{tabular}{ccccc}
    \toprule
    Method &  SimCLR~\cite{zhang2021contrastive} & DINO & LGL\cite{tao2021self} & DLG-LC   \\
    \midrule
    Loss-Gate & $\times$ & $\times$ & \checkmark & \checkmark \\
    \midrule
    Iter-1 & 6.281	& 4.255	& 3.520	& \textbf{2.723} \\
    Iter-2 & 5.914	& 3.946	& 2.410	& \textbf{1.888} \\
    Iter-3 & 5.547	& 3.824	& 2.070	& \textbf{1.670} \\
    Iter-4 & 4.872	& 3.691	& 1.950	& \textbf{1.495} \\
    Iter-5 & 4.484	& 3.510	& 1.660	& \textbf{1.468} \\
    \bottomrule
  \end{tabular}
  \vspace{-0.4cm}
\end{table}


Then, we summarize the performance in each iteration with DLG-LC on Vox-O set and the results are presented in Table~\ref{tab:DLGLC_iteration}. We first compare the iterative results of generating pseudo labels using DINO and SimCLR respectively. And they are both trained on all training data without any loss-gate strategies.
Obviously, DINO owns consistent advantages over SimCLR in each iteration which demonstrates DINO can produce more discriminative speaker embedding for generating more reliable pseudo labels. 
After applying the loss-gate policy, results have been significantly improved. Our proposed DLG-LC with dynamic threshold shows great advantages compared with LGL. It not only has superior performance but also has a faster convergence speed. Only with 3 iterations, DLG-LC has achieved comparable results with the final iteration of LGL, which is benefit from the dynamic threshold and label correction. 

Finally, a performance comparison with other self-supervised speaker verification systems and our proposed DLG-LC is given in Table~\ref{tab:DLGLC_overview}. Most of them are from the latest Voxceleb speaker recognition challenge (VoxSRC)~\cite{nagrani2020voxsrc,brown2022voxsrc} which represent the most advanced performance nowadays. From the table, it's obviously observed that DLG-LC exceeds all the existing methods and outperforms the best one by relative $11.45\%$, $18.35\%$ and $15.16\%$ on Vox-O, Vox-E  and Vox-H sets respectively with only 5 iterations. Moreover, the clustering algorithm we adopted is $k$-means which is simpler than agglomerative hierarchical clustering (AHC) used in other systems. Besides, we employ 7,500 instead of 6,000 as the number of clusters because 5,994 is the real number of speakers in the training set. Nevertheless, our system still achieves the \textbf{state-of-the-art} performance, which also shows the robustness of our system.

\begin{table}[h]
  \caption{EER (\%) comparison for different self-supervised speaker verification methods on Vox-O, E, H}
  \label{tab:DLGLC_overview}
  \centering
  \begin{tabular}{lccccc}
    \toprule
    Method &  Clustering & Iter & Vox-O & Vox-E & Vox-H   \\
    \midrule 
ID\cite{idlab2020} &	AHC(7500) &	7&	2.10 &	-&	- \\
JHU\cite{jhu2021}&	AHC(7500) &	5&	1.89 &	-&	- \\
DKU\cite{dku2021}&	K-M(6000) &	4&	1.81 &	-&	- \\
SNU\cite{SNU2021}&	AHC(7500) &	5&	1.66 &	-&	- \\
LGL\cite{tao2021self}& K-M(6000) &	5&	1.66 &	2.18&	3.76 \\
DLG-LC & K-M(7500) &	5&	\textbf{1.47} &	\textbf{1.78} &	\textbf{3.19} \\
    \bottomrule
  \end{tabular}
  \vspace{-0.4cm}
\end{table}

\section{Conclusions}
In this work, we propose dynamic loss-gate and label correction (DLG-LC) to alleviate the performance degradation of self-supervised speaker verification caused by inaccurate assumptions and labels. By modeling the loss histogram with Gaussian distribution, we can obtain a dynamic loss-gate to select reliable data when training on pseudo labels. Instead of dropping unreliable data, we incorporate a predicted posterior probability as the target distribution to prevent fitting into incorrect samples. The experiments on Voxceleb show that our proposed DLG-LC is more robust and achieves the \textbf{state-of-the-art} performance.


\section{Acknowledgements}
This work was supported in part by China NSFC projects under Grants 62122050 and 62071288, and in part by Shanghai Municipal Science and Technology Major Project under Grant 2021SHZDZX0102. Experiments have been carried out on the PI super-computer at Shanghai Jiao Tong University. 

\bibliographystyle{IEEEtran}

\bibliography{mybib}

\end{document}